# Correlation-driven metal-insulator transition in unconventional magnetic metal superoxides

Sarajit Biswas,[1,2] Pratim Banerjee[2] and Molly De Raychaudhury[2*]

[1]Department of Physics, Barasat Government College, Kolkata-700124, West Bengal, India
[2]Department of Physics, West Bengal State University, Kolkata-700126, West Bengal, India

*Corresponding author's email: *molly@wbsu.ac.in*

**Abstract**

Using first-principles electronic structure calculations, we have extensively studied the electronic and magnetic properties of alkali sodium superoxide ($NaO_2$) in comparison with that of potassium superoxide ($KO_2$) both at high and low temperatures. These properties of these superoxides are governed by the unpaired electron donated by the alkali atoms Na and K to the O atoms forming dimers. This unpaired electron is the source of orbital fluctuations in the O-$\pi^*$ manifold for both cases. In order to reduce this orbital fluctuation, both go through several structural phase transitions. In these plethora of structures, the $O_2^-$ dimers undergo rotation, leading to a complex linking of its orbital degrees of freedom with its spin degrees of freedom. Hence the magnetic properties are found to be controlled by this unpaired electron vary as the orientations of these $O_2^-$ dimers change. Due to the change in the orientations of $O_2^-$ dimers, the alkali ion cages around the $O_2^-$ dimers change from square in the pyrite phase to rhombus and rectangle for the orthorhombic phase for $NaO_2$ and square in the tetragonal phase to parallelogram in the monoclinic phase for $KO_2$ on the plane cutting through the dimers. The band structures of $NaO_2$ in the low-temperature orthorhombic phase and $KO_2$ in the monoclinic phase show that the lifting of degeneracy in the O-$\pi^*$ manifold is due to the redefined electrostatic interaction between the K/Na cages and the $O_2^-$ dimers. This, in addition to electron correlation among the localized O-$\pi^*$ electrons, establishes complete orbital ordering (OO) in turn drives metal-insulator transition (MIT) in both the systems. Furthermore, K doping for Na in $NaO_2$ also results in correlation-induced MIT, predicted to take place at a temperature higher than that in $NaO_2$. This opens up the possibility of MIT in Rb/Cs-doped $NaO_2$ at even higher temperatures.

# 1. Introduction

Alkali RO$_2$ superoxides (where R = Na, K, Rb, Cs) exhibit complex magnetic phase diagrams featuring the interplay of spin, orbital, and lattice degrees of freedom. They are made up of alkali cations R$^+$ and superoxide O$_2^-$ anions, which are charged oxygen molecules that have gained an extra electron from R. The superoxide anion (O$_2^-$) is a fascinating entity, playing important physiological functions in biological processes [1]. In recent years, NaO$_2$ has received a lot of interest because of its unconventional source of dipole moment, complex magnetic phase diagram arising out of the magnetogyration effect and the primary discharge product in a Na-air battery [2-29]. This material is unique in the sense that the p-electrons generate magnetism in this material. On the other hand, conventional magnetic materials are often made up of transition metals and rare-earth elements, with *d* and *f* electrons giving magnetism, respectively. As a result, NaO$_2$ can be used to replace conventional magnetic materials, not found in abundance in nature. Therefore, this material can be considered to be a feasible alternative to limited magnetic transition metal or rare-earth compounds, and has a wide range of applications, including the fabrication of oxygen regeneration devices [16, 22-26]. The most important oxide formed by Na is NaO$_2$ in which each Na and O have oxidation states of +1 and -0.5 respectively. Therefore, each O$_2^-$ complex has an extra electron, which is acquired from the Na atom. The extra electron remains unpaired in O$_2^-$ complex, resulting in paramagnetism in NaO$_2$ at high temperature. The superoxide group (O$_2^-$) acts as the anion and exhibits a variety of properties that are assumed to be related to superoxide ion properties [4, 12, 16, 29].

The first investigation on NaO$_2$ was reported by Templeton and Daubel in 1950 [2]. At room temperature (RT), cubic NaO$_2$ with Fm-3m crystal symmetry, is paramagnetic with an orientational disorder among O$_2$ entities. However, all O$_2^-$ dimers are oriented randomly in this phase [7]. Upon decreasing the temperature, the Fm-3m structure transforms into rock salt-type pyrite (Pa-3) structure at around 230 K and finally assumes the orthorhombic $FeS_2$-type (Pnnm) structure in the temperature range of 196 K to 43 K. This phase is commonly known as the marcasite phase. At 230 K, $O_2^-$ dimers orient themselves in four directions and hence exhibit paramagnetism with long-range orientational order. At temperatures below 196 K, the $O_2^-$ dimers, restrict themselves to two directions of orientation and hence are parallel in alternative planes. The main variation among the three structural phases is the $O_2^-$ orientational order, which possibly results in its complex magnetic behaviour. The relative orientation of the $O_2^-$ complexes and the state of the partially occupied O-$\pi^*$ orbitals appear to be governing the magnetic behaviour of $NaO_2$ [8, 12, 16, 29].

Even though there are quite a few reports on the electronic properties and magnetic ground states of $NaO_2$, Solovyev *et al.* [16] and Biswas *et al.* [8] have thoroughly investigated these properties and presented a detailed scenario of the electronic properties and magnetic ground states of various phases of $NaO_2$. But it is not clear as to why this system goes through the various orientations of $O_2^-$ dimers. The role of the alkali cation Na is also not unambiguously established. High symmetry and thus degeneracy among the O-2p orbitals are electronic features of the cubic phase. At ambient temperature, the freely rotating $O_2^-$ molecules (weak dimers) are the source of permanent dipole moment, which leads to paramagnetism in cubic $NaO_2$. High symmetry causes the O-$2p_x$, $2p_y$ and $2p_z$ atomic orbitals to be degenerate. The same orbitals in $O_2^-$ form $\sigma$ and $\pi$ bonding and antibonding orbitals, maintaining their degeneracy and are band-like in the cubic phases. In the pyrite phase, the Na-O bond weakens and $O_2^-$ dimers become stronger. Even then, the dimers are now constrained to rotate in four directions [see Fig. 1]. As a result, the degeneracy between the $\sigma$ and $\pi$ manifolds is lifted to some extent, but the cubic structure still causes retention of some order of degeneracy between the $\pi$ orbitals. The <O-Na-O angle is 180° along the c-axis, which favours superexchange-mediated antiferromagnetism (AFM). In the ab plane, the <O-Na-O angle is 90°, fevouring ferromagnetism (FM) due to the superexchange mechanism. As a result, competition between FM and AFM interactions is observed in the pyrite phase [16]. Although the orbital fluctuation is decreased to some extent, the orbital fluctuations in order to fill one of the two degenerate $\pi^*$ orbitals with the single electron transferred from Na cation possibly leads to the structural transition from Fm-3m to Pa-3. This leads to a redefining of the delocalized O-2p bands of the Fm-3m phase into pure molecular orbitals, namely $\sigma$, $\pi$, $\sigma^*$ and $\pi^*$.

The nine electrons on $O_2^-$ dimers lead to the electronic configuration of $\sigma^2$, $\pi^4$, $\pi^{*3}$ and $\sigma^{*0}$. The empty $\sigma^*$ manifold is on account of the strong direct overlap of O-$2p_z$ orbitals of the strong dimers. In this phase, the degeneracy between O-$p_z$ and $p_{x/y}$ atomic orbitals is broken. The $p_x$ and $p_y$ orbitals combine to form the $\pi$ and $\pi^*$ orbitals, whereas $p_z$ orbitals overlap to form $\sigma$ orbitals. The degeneracy between the two $\pi^*$ orbitals, is, however not broken and hence the $\pi^*$ orbital is occupied in the spin majority channel and the degenerate $\pi^*$ orbital is partially filled in the spin minority channel. Therefore, orbital fluctuation is sustained in the $\pi^*$ orbital in one spin channel but is at a reduced scale in the pyrite phase. In the marcasite phase, the $O_2^-$ dimers are aligned parallel in alternate planes [Fig. 1]. The $\sigma$ and $\pi$ molecular orbitals are completely separated from each other. In addition, the transition to the marcasite phase relieves the degeneracy among the $\pi^*$ orbitals to some extent [8], and the localization of O-2p orbitals is enhanced significantly compared to the pyrite phase with the dimer length increasing markedly. The maintenance of orbital fluctuations is possibly the reason behind the degeneracy among the two O-$\pi^*$ orbitals of $NaO_2$. Biswas *et al.* [8] pointed out that a superexchange mediated AFM ground state in $NaO_2$ might become possible if the orbital fluctuation is completely removed as in $KO_2$ [29, 34].

The K-counterpart, $KO_2$ is also an attractive superoxide due to its magnetogyration-driven AFM ground state and controversial electronic properties [3, 4, 29-42]. At high temperatures, this compound has a cubic NaCl structure with disoriented oxygen molecules [37]. Upon cooling, it encounters a structural phase transition (SPT) from the cubic to tetragonal I4/mmm phase at 395 K [4, 30] with an average $O_2$ orientation parallel to the tetragonal axis. Below 231 K, an incommensurate phase is observed, and the crystal structure stabilizes in the monoclinic C2/c phase [31]. At lower temperatures, the symmetry is further lowered to the triclinic phase, where neutron scattering experiment [32] discovered a long-range A-type antiferromagnetic (AFM) order with opposite spin directions in two adjacent ferromagnetic (00l) oxygen layers below the Néel temperature $T_N$ = 7 K. The electronic properties of the $KO_2$ crystal are tightly linked to the orbital structure of the $O_2^-$ ions, in which the hybridization between atomic 2p orbitals gives rise to molecular bonding σ, $π_+$, $π_-$ and antibonding $σ^*$, $π_+^*$, $π_-^*$ levels [31]. In presence of Coulomb interaction U, the monoclinic phase exhibits an insulating gap due to the reduction of symmetry and, thereby the lifting of degeneracy and the commencement of orbital ordering (OO) [29, 31, 33]. Sikora *et al.* [31] reported that an application of Coulomb interaction of U=4 eV on the O-2p electrons removes the orbital degeneracy in the monoclinic phase and the system thereby encounters MIT. They also have shown that the AFM insulating ground state is necessarily driven by correlation but not essentially by spin-orbital coupling if ferro-orbital coupling ordering is triggered. We have tried to understand the effect of orbital ordering as the $O_2^-$ dimers undergo rotation with decreasing temperature in $KO_2$ as compared to that in $NaO_2$.

We have applied Hubbard-type on-site Coulomb interaction U (in the range of 1 eV to 6 eV) on the O-2p electrons first in the pyrite phase and then also for the marcasite phase of $NaO_2$, in order to introduce the correlations among the localized O-2p molecular orbitals. Besides, for the low-temperature monoclinic phase of $KO_2$, the application of the same U (=4 eV), as applied by Sikora *et al.* [31], was found to be responsible for the MIT in $KO_2$. We have also carried out a comparative study of the electronic and magnetic properties of $KO_2$ with an aim to gain full insight into the unique phenomenon of magnetogyration-assisted MIT in superoxides. Finally, we have doped 50% of Na atom with K in $NaO_2$ (the corresponding compound is $Na_{0.5}K_{0.5}O_2$) and explored the role of the alkali metal in driving the system into an insulating state.

## 2. Computational methods

In the present study, the electronic and magnetic properties of $NaO_2$ in its marcasite phase were calculated utilizing density functional theory (DFT) [44, 45], as implemented in the linearized-muffin-tin orbitals (LMTO) method [46, 47]. For the exchange-correlation interactions, the local spin density approximations (LSDA) [47, 48] approach was used. We also employed LSDA calculations with the inclusion of Hubbard-type on-site Coulomb interaction U (LSDA+U) [49, 50] method for the electronic and magnetic properties of pure $NaO_2$ and $KO_2$ in their low-temperature phases and later for the K-doped compound $Na_{0.5}K_{0.5}O_2$. A total of 6x6x8 = 288 k-points were considered in the Brillouin zone (BZ) to calculate band structure. For structural relaxation, the Vienna ab initio simulation package (VASP) [51] was employed. The k-mesh used to perform BZ integration is 12×8×16=1536. The energy cutoff, energy criterion for ionic relaxation and electronic self-convergency criterion were used as 400 eV, 0.1 eV and $10^{-4}$ eV respectively. The conjugate gradient algorithm with scaling parameter 0.5 was also used for structural relaxation. The lattice parameters used for the marcasite $NaO_2$ at 77 K are a = 4.332 Å, b = 5.54 Å and c = 3.364 Å [43]. Two equivalent Na and four equivalent O atoms make up the unit cell. The unit cell is increased to a supercell in such a manner that it contains four Na and eight O atoms (the corresponding compound is $Na_4O_8$) i.e. four formula units (f.u.) per unit cell. We substituted two Na atoms with two K atoms to attain 50% doping level (the corresponding compound is $Na_2K_2O_8$). For the monoclinic phase of $KO_2$, the lattice parameters used are a=8.098 Å, b=4.2 Å, c=8.178 Å and β = 125.093°. The unit cell consists of two equivalent K and four equivalent O-atoms. We have also studied the same electronic and magnetic properties of $NaO_2$ within LSDA+U formalism for U lying in the range of 0 to 6 eV on O-2p electrons. It is to be noted that several pieces of evidence are available for applying high U on the O-2p electrons [31, 34, 52-54]. For example, Jing *et al.* [52] applied U=6 eV on the O-2p electrons for correcting the band gap underestimation in $BaBiO_3$. The calculated band gap of 1.88 eV for the $BaBiO_3$ system is close to the experimentally measured optical band gap of 2.05 eV [53]. Kulik *et al.* [54] also applied a high U value of 8.5 eV on the O-2p electrons. For $KO_2$ system, Sikora et al. [31] used U=4 eV on the O-2p electrons. Nandy et al. [34] also used U in the range of 3 to 6 eV on the O-2p electrons for $KO_2$. Hence, LSDA+U method was applied to all three NaO$_2$, KO$_2$ and 50% K-doped NaO$_2$ (the corresponding compound is Na$_{0.5}$Ka$_{0.5}$O$_2$) to investigate their electronic and magnetic properties.

## 3. Results and discussion

The supercell of marcasite NaO$_2$ has four Na and eight O atoms i.e. four formula units. The dimers are found to orient in two directions [Fig. 1(a)]. The orientations of the two O$_2^-$ dimers are also shown in Fig. 1. The dimers are arranged parallelly in alternate planes. All the O-O distances in the O$_2^-$ dimer are uniform =1.31 Å (usually molecular oxygen bond lengths =1.21 Å). The Na-coordination number of each O$_2^-$ dimer is six, which forms a O$_2$Na$_6$ octahedron. Four of these Na atoms are found almost in a plane that cuts the oxygen dimer into halves forming the basal plane, while the other two are found along the line (apical plane) that connects two oxygen atoms of the same kind. The O$_2^-$ dimers with two identical orientations are arranged in alternate ab planes. Consequently, the orientational disorder is reduced even more than in the pyrite phase [8]. All the Na-O distances are found uniform with a magnitude of 2.3814 Å. The O atoms tend to move away from each other after gaining an extra electron from Na atom. As a result, an increase in the dimer lengths is observed in the structure of NaO$_2$ [8]. However, after a certain point, the presence of Na atoms and hence the electrostatic repulsion between the Na and O electrons restricts this movement. Consequently, O$_2^-$ dimers are rotated from their axial direction. This effect is known as the 'magnetogyration' effect. The magnetogyration effect is observed simultaneously with different Na-O bond distances and <Na-O-Na angles, as reported in Table 1.

More interestingly, we observed two types of Na cages around O$_2^-$ dimers. One cage is rhombus-shaped, while the other cage is rectangular-shaped. The formation of these two types of alkali metal cages is schematically illustrated in Fig. 1 (b). The uniform Na-Na side distances in the rhombus-type Na cage is 3.8978 Å, whereas one of the opposite pair of <Na-Na-Na angles are 89.425° and 90.574°. On the other hand, Na-Na side distances are 3.364 Å and 4.3319 Å for the rectangular Na cage with all <Na-Na-Na=90°. The Na-Na diagonal distances for the rhombus-shaped Na cage are 5.54 Å and 5.4755 Å, whereas the Na-Na diagonal distance for the rectangular-shaped Na cage is 5.4848 Å. In comparison, we observed in the pyrite phase [crystal structure and the formation of $O_2^-$ dimers are illustrated in Fig. 1(c)], only one kind of Na cage having square shape [as illustrated in Fig. 1(d)]. In the Na cage, Na-Na side lengths are uniform with a magnitude of 3.8749 Å and all <Na-Na-Na angles are 90°, and the Na-Na diagonal distance is 5.48Å.

Therefore, Na-Na distances remain almost unmodified (only increased by 0.02 Å) for the square-shaped Na cage, whereas it is reduced/enhanced by 0.5109 Å/0.457 Å for the rhombus-shaped Na cage due to the SPT from the pyrite to marcasite phase. In addition, <Na-Na-Na angles are reduced/enhanced by 0.574°/0.425° for the rhombus-shaped Na cage and remain unaltered for the rectangular-shaped Na cage. Also, Na-Na diagonal distances are enhanced/reduced by 0.06 Å/0.045 Å for the rhombus-shaped Na cage, whereas the same for the rectangular-shaped Na cage remains almost unmodified (enhanced only by 0.0048 Å). As a result, the orthorhombic distortion (where a ≠ b ≠ c) in the marcasite phase renders the alkali cage a significantly higher distortion compared to that in the cubic pyrite phase (where a=b=c). These enhanced distortions in the Na cage are associated with a transition from orientational disorder to partial order of the $O_2^-$ dimers. This structural aspect is also related to the simultaneous development of short-range magnetic order but has unfortunately not been characterized with unambiguity [4]. This is the magnetogyration effect or modification of the orientation order of the dimers, which should significantly modify the electronic structure of marcasite $NaO_2$ compared to its pyrite phase.

The total and partial band structures (PBS) of $NaO_2$ in the pyrite phase are illustrated respectively in Figs. 2(a) and 2(b) for the spin majority channel and in Figs. 2(c) and 2(d) for the spin minority channel. In the majority spin channel, O-σ, π, $π^*$ and $σ^*$ states are mainly found between -10.48 to -9.36 eV, -9.36 to -8.2 eV, -2.25 to -0.4 eV and -2.25 to 6.0 eV respectively. The corresponding orbitals for the spin minority channel are mainly found between -9.7 to -8.4 eV, -8.4 to -7.3 eV, -1.6 to 0.7 eV and -1.6 to 6.0 eV respectively. We found that all the atomic orbitals, O-$p_x$, $p_y$ and $p_z$ forming the molecular O-σ, π, $π^*$ and $σ^*$ states are degenerate for both spin channels [Fig. 2(b)/Fig. 2(d) for the spin majority/minority channel]. A specific combination of the antibonding components of $p_x$ and $p_y$ orbitals forms $\sigma^*$ orbitals (which are found above the $E_F$) and the other antisymmetric combination of the antibonding components of $p_x$, $p_y$ and $p_z$ orbitals forms $\pi^*$ orbitals. Thus, orbital fluctuation is anticipated when the Na atom transfers its 3s electron and has to be filled in one of these degenerate bands. Hence we find orbital degeneracy leading to the metallic behaviour in this system.

Figure 3 depicts the spin-polarized total density of states (TDOS) and atom-, l-projected density of states (PDOS) of pyrite $NaO_2$ for U = 0. The system is metallic with low carrier concentration in the majority spin channel and fully metallic in the minority spin channel. It is evident that the TDOS is dominated by the O-2p character. The calculated magnetic moments on Na-3s and O-2p electrons are -0.19 $\mu_B$ and -0.49 $\mu_B$ respectively. Therefore, the electronic and magnetic properties of $NaO_2$ are primarily determined by the O-2p electrons. It is also observed from Fig. 3 that states due to the $O_2^-$ dimers are similar to molecular $O_2$ states, namely the fully occupied σ and π bonding states and the partially occupied π antibonding ($\pi^*$) and fully unoccupied σ antibonding states ($\sigma^*$). It can be concluded that the O-σ and π orbitals are fully occupied in both the spin channels, whereas $\pi^*$ orbitals are fully occupied in the majority spin channel and are partially occupied in the minority spin channel. The $\sigma^*$ orbitals are fully unoccupied in both the spin channels. The single electron from Na-3s orbital is gained by the dimers (of length =1.158 Å). Hence, Na-3s orbital is unoccupied. However, this single electron occupies the partially filled degenerate O-p manifolds as stated earlier too. This competition among the six O-p orbitals of the $O_2^-$ dimers to accommodate the single electron introduces quantum fluctuations in the system.

Also interestingly, we found that the π and $\pi^*$ states form narrow bands, indicating the localized nature of some of the O-2p electrons. Thus, to see the effect of electronic correlations among the O-2p electrons on the electronic structure of the present system, we have applied U in the range of 1 to 6 eV on the O-2p electrons. Throughout this range of U, the electronic structure does not change significantly, and the system preserves its metallicity. Figure 4 shows the TDOS and PDOS of pyrite $NaO_2$ for U = 6 eV. The only modification that occurs is the localized parts of

$\pi/\pi^*$ orbitals become more localized. However, this does not lift the degeneracy among the O-2p orbitals, sustaining the metallicity and also the orbital fluctuation in the system.

Now, this orbital fluctuation would drive the system to assume a structure of lower symmetry, namely the orthorhombic structure in the marcasite phase in this case. The total and partial band structures of O-$p_{x/y}$ and $p_z$ orbitals for the spin majority channel are displayed respectively in Figs. 5(a), 5(b) and 5(c). The corresponding band structures for the spin minority channel are illustrated respectively in Figs. 5(d), 5(e) and 5(f). We observed that contrary to that in the pyrite phase, the O-$p_x$ and $p_y$ orbitals are exactly degenerate, whereas $p_z$ orbitals are non-degenerate in both the spin channels in the marcasite phase. In the spin majority channel, the O-$\sigma$, $\pi$, $\pi^*$ and $\sigma^*$ orbitals are found around -7.2 to -6.48, -6.48 to -4.88 eV, -1.85 to -0.35 eV and 1.7 to 4.4 eV. In the spin minority channel, these orbitals are found around -6.57 to -5.86 eV, -5.86 to -4.25 eV, -1.32 to 0.38 eV and 2.0 to 4.6 eV respectively. **The third manifolds consist of a very localized part and a small delocalized part.** It is evident from these figures that the $\sigma$ molecular orbitals originated from a specific combination of $p_x$, $p_y$ and $p_z$ atomic orbitals, whereas the $\pi$ molecular orbitals originated from a different combination of $p_x$, $p_y$ and $p_z$ atomic orbitals. Clearly, there is a lifting of degeneracy among the O-2p orbitals. The separation of $\sigma/\sigma^*$ orbitals from the $\pi/\pi^*$ manifold is sufficiently evident but the lifting of degeneracy among the $\pi/\pi^*$ orbitals is not very apparent. Nevertheless, the filling of O-2p orbitals is shown in Fig. 6. The $\pi/\pi^*$ orbitals have a localized component too.

The spin-polarized TDOS and PDOS of $NaO_2$ for this phase with U=0 eV as shown in Fig. 7 implies that the orthorhombic distortion drives the system to a half-metallic state. The contributions to the TDOS from the Na-3s (green lines) and Na-3p (blue lines) electrons are insignificant as compared to the O-2p electrons (red lines). The calculated magnetic moments on Na-3s and O-2p electrons are -0.14 $\mu_B$ and -0.42 $\mu_B$ respectively. Thus, the electronic and magnetic properties of $NaO_2$ are here too primarily determined by the O-2p electrons. It is to be mentioned that the dimer length has now become to 1.31 Å. And what is noteworthy is that the basal Na cage, which was square in the pyrite phase, is now of two types: rectangular-shaped and rhombus-shaped. The orthorhombic distortion, where $a \neq b \neq c$ renders such distortion in the alkali cage. The profound effect of this distortion on the electronic structure is the lifting of degeneracy in the $\pi/\pi^*$ manifolds.

Strong correlation among some of the O-2p electrons comprising the $\pi/\pi^*$ orbitals is now included via the LSDA+U formalism in the marcasite phase. The TDOS of $NaO_2$ for different U values in the range of 3 to 6 eV are shown in Fig. 8. The system preserves its half-metallicity up to U = 5 eV, and eventually encounters MIT at U=6 eV. As stated earlier, the $O_2^-$ dimer length is increased to 1.31 Å from 1.158 Å, as the structure changes from pyrite to marcasite. In the former, two oxygen atoms exist as if they are part of molecular oxygen in an alkali cage. The latter is a weaker dimer but strongly interacting one with the Na cage. As a result, the distortion in the Na cage is expected to be higher as the system tries to overcome orbital fluctuation.

The σ and π states are occupied in an energy window of -7.0 eV to -2.7 eV of width 4.3 eV. The separation between the bonding and antibonding π orbitals is about 8 eV and that of σ orbitals is very large ~13 eV. The total, O-$p_{x/y}$ and $p_z$ partial band structures of $NaO_2$ for U =6 eV are depicted in Figs. 9(a), 9(b), 9(c), and Figs. 9(d), 9(e), 9(f) for the spin majority and spin minority channels, respectively. We observed a marked modification in the band structure of O-$\pi^*$ orbitals for both the spin channels. But a significant modification occurs in the minority spin channel. When U is applied, the σ and $σ^*$ states remain unaffected because they are delocalized. Correlation among the O-2p electrons contributing to the localized $\pi/\pi^*$ manifolds lifts the degeneracy among these orbitals completely. The SPT from the pyrite to marcasite phase not only separates or removes the σ from the π manifold but also lowers the symmetry in such a way that the π and $\pi^*$ manifolds are split into $\pi_+$, $\pi_-$ and $\pi_+^*$, $\pi_-^*$ states respectively. Each of σ, $\pi_+$ and $\pi_-$ states are now occupied by two O-2p electrons each and hence are fully occupied. The orthorhombic distortion deforms the Na cage in such a way that the degeneracy in the $\pi_+^*$ and $\pi_-^*$ states is lifted, with the former becoming fully occupied and the latter being half-filled. This introduces orbital ordering (OO) in the system by reducing the orbital fluctuation among the $\pi^*$ states. The system gets stabilized by getting the unpaired single electron donated by Na filled in the $\pi_+^*$ states. The splitting and filling of O-2p orbitals of $O_2^-$ dimers here is schematically illustrated in Fig. 10. The occupied levels are shifted below the $E_F$ and the unoccupied levels are shifted above the $E_F$ by application of U, leading to a complete OO and hence the opening of a gap of 0.3 eV occurs near the $E_F$.

The TDOS and PDOS of Na-3s, Na-3p and O-2p orbitals of pure $NaO_2$ are displayed in Fig. 11. We observed significant modifications in the electronic structure for the O-π and $\pi^*$ orbitals, especially for the spin minority channel. The separation of π orbitals into $\pi_+$ (-6.35 to -5.63 eV/-5.25 to -4.5 eV for the spin majority/minority channel) and $\pi_-$ (-5.63 to -5.25 eV/-4.5 to -2.65 eV for the spin majority/minority channel) orbitals is more pronounced for both the spin channels. We observed the tendency of $\pi^*$ orbitals to get separated into $\pi_+^*$ and $\pi_-^*$ orbitals in the spin majority channel (-2.25 to -0.9 eV). So we find that if the degeneracy in the O-$\pi^*$ orbitals could be lifted completely, full OO could be established, leading to metal-insulator transition (MIT) in $NaO_2$. Therefore, correlation-induced complete OO triggered by structural distortion in the Na cage can drive MIT in $NaO_2$. It is anticipated that full OO should facilitate antiferromagnetism in $NaO_2$ as in $KO_2$. This aspect we shall investigate later, as experimental data is yet inconclusive. Let us now try to understand the magnetogyration phenomenon in $NaO_2$ as compared to that in $KO_2$.

The crystal structure of monoclinic $KO_2$ is displayed in Fig. 12(a). The orientation of $O_2^-$ dimers is also shown in both Figs. 12(a) and 12(b). The $O_2^-$ dimers are also oriented parallel in alternative planes as is observed in marcasite $NaO_2$. The O-O distances are identical (=1.3534 Å) for both $O_2^-$ dimers. Furthermore, each $O_2^-$ dimer has the coordination of six K atoms, forming an octahedron around the $O_2^-$ dimer (in a similar manner as in the case of $NaO_2$). The K-O distances and <K-O-K angles for the $O_2K_6$ octahedra are also reported in Table 1. We observed only one kind of K cage around $O_2^-$ dimers, which is parallelogram-shaped and is schematically depicted in Fig. 12 (b). The K-K side distances in the cage are 4.1411 Å and 4.2 Å, whereas the <K-O-K angles are 80.896° and 99.103°. The K-K diagonal distance is 5.8982 Å. To compare the degree of structural distortion in the monoclinic phase, we also investigated the crystal structure of $KO_2$ with the tetragonal structure [shown in Fig. 12(c)]. In this structure, we also observed only one type of K cage, i.e. square-shaped. The formation of square-shaped K cage is illustrated in Fig. 12(d). The uniform side K-K distance in the K cage is 4.034 Å. The K-K diagonal distance is 5.7049 Å. The average depletion in side/diagonal K-K distances in the square-shaped K cage is 0.88 Å/0.19 Å. On the other hand, <K-K-K varies significantly by an amount ~9.1°. Due to the SPT from the pyrite to marcasite phase, the average depletion in both side and diagonal Na-Na distances in the rhombus-shaped Na cage is ~0.02 Å. For the rhombus-shaped Na cage, the average side Na distance is increased by ~0.05 Å and the diagonal Na distance remains almost unmodified (increases only by ~0.005 Å. The <Na-Na-Na increases only by an amount of ~0.6° for the rhombus-shaped Na cage. Therefore, the lowering of crystal symmetry in monoclinic $KO_2$ (SPT from the tetragonal phase) is much greater compared to the lowering of symmetry about the SPT in $NaO_2$ from the pyrite to the marcasite phase. The enhanced lowering of crystal symmetry in monoclinic $KO_2$ can be anticipated from the fact that the more loosely bound and extended 4s electrons of K ions yield stronger electrostatic interaction on O-2p electrons than that exerted by 3s electrons of Na ions in marcasite $NaO_2$. This results in a stronger magnetogyration effect in $KO_2$. This is not surprising as the $O_2^-$ dimers are oriented in two directions comparatively at a much higher temperature in $KO_2$ as compared to the dimer achieving two orientations at the best at a much lower temperature in $NaO_2$, it is expected that the doping with K at the Na site will have a dramatic effect on the MIT in $NaO_2$.

The total and partial band structures of O-$p_x$, $p_x$, $p_z$ orbitals of monoclinic $KO_2$ for the spin majority channel are displayed in Figs. 13(a), 13(b), 13(c) and 13(d) respectively. For the spin minority channel, the corresponding band structures are illustrated respectively in Figs. 13(e), 13(f), 13(g) and 13(h). In the spin majority channel, O-σ, π, $π^*$ and $σ^*$ orbitals are mainly found around -6.86 to -6.0 eV, -6.0 to -4.5 eV, -1.25 to -0.1 eV and 4.5 to 8.0 eV respectively. For the spin minority channel, the corresponding orbitals are found mainly around -6.2 to -5.2 eV, -5.2 to -3.8 eV, -0.6 to 0.7 eV and 5.0 to 8.0 eV respectively. Unlike marcasite $NaO_2$, where the $p_z$ orbital was not degenerate with the other two orbitals, here the O-$p_x$ and $p_y$ orbitals are also not degenerate, indicating much lowering of structural symmetry in $KO_2$. The $\sigma/\sigma^*$ and $\pi/\pi^*$ orbitals are found to be separated from each other for both spin channels [see Figs. 13(b), 13(c), 13(d) and Figs. 13(f), 13(g), 13(h)]. This event is also observable from the LSDA TDOS of monoclinic $KO_2$, as depicted in Fig. 14. It is worth mentioning that the system is half-metallic with a pseudogap at $E_F$ in the conducting channel [see Fig. 14].

More importantly, the degeneracy among the $\pi/\pi^*$ orbitals is lifted. This has a profound effect on the two $\pi^*$ orbitals around the $E_F$, where the $\pi^*_+$ orbital becomes occupied and the other $\pi^*_-$ orbital is shifted above the $E_F$. This orbital degeneracy lifting at the $E_F$ becomes more evident by applying Hubbard U as should be applied for the correlated O-$\pi/\pi^*$ electrons. The effect of correlations on the O-2p electrons is shown in the spin-polarized TDOS and PDOS of $KO_2$ within LSDA+U, as illustrated in Fig. 15. We applied U in the range of 1 to 4 eV and found that $KO_2$ encounters MIT for U=4 eV. With the application of U, we find that a small gap of 0.3 eV is introduced in the minority spin channel, while the majority spin channel remains insulating. The magnetic properties of $NaO_2$ measured using neutron scattering is inconclusive about the magnetic ground state [4], whereas $KO_2$ is known to be an A-type antiferromagnet below 7 K [32]. The latter undergoes an SPT from its monoclinic to triclinic phase at 12 K [4]. But we have observed that $KO_2$ undergoes an MIT as it assumes a monoclinic structure at 197 K, implying that the electronic and the magnetic properties might not be related to each other. Hence, taking a cue, we tried to drive an MIT in $NaO_2$ by tuning the nature and concentration of *s*-electrons, namely by doping at the Na-site by K ions. By doing so, our aim was to make the role played by the s-electrons of alkali ions clearer. In this direction, we have investigated the effect of the alkali metal in causing the system to become insulating by replacing 25%, 50% and 75% of the Na atoms in marcasite $NaO_2$ with K atoms.

After structural relaxation of a unit cell formed by substituting K at Na site in $NaO_2$, $Na_{0.5}K_{0.5}O_2$ was formed. $Na_{0.5}K_{0.5}O_2$ adopts the monoclinic structure with lattice parameters a =8.140 Å, b =7.471 Å, c=9.361 Å and β =88.07°. Therefore, $NaO_2$ exhibits an SPT from the marcasite phase to the monoclinic phase upon 50% K doping for Na. The crystal structure of monoclinic $Na_{0.5}K_{0.5}O_2$ is shown in Fig. 16 (a). In Fig. 16(b), the orientations of $O_2^-$ dimers are also depicted. It is observed that the $O_2^-$ dimers orient parallel in alternative planes. Interestingly, we found two different types $O_2^-$ dimers. The red-coloured $O_2^-$ dimers, or Dimer-I, have an O-O distance of 1.3285 Å, whereas the orange-coloured $O_2^-$ dimers, or Dimer-II, have an O-O distance of 1.3686 Å. Therefore, the mean $O_2^-$ dimer distance is 1.3486 Å. Consequently, as compared to pure $NaO_2$, the $O_2^-$ dimer distance in $Na_{0.5}K_{0.5}O_2$ is increased modestly by an amount of 0.0386 Å. Three each of Na and K atoms coordinate to form a Na/K octahedron ($O_2Na_6$ or $O_2K_6$) around each $O_2^-$ dimer. Table 1 also lists the Na/K-O distances and <Na/K-O-Na/K angles for the Na/K octahedra. Around the two $O_2^-$ dimers, two different kinds of alkali cages were observed. The alkali cage [schematically depicted in Fig. 16(b)] that surrounds the red-coloured dimers (Dimer-I) is rectangular in shape and is made up of four K atoms. The K-K side distances in the cage are 3.9534 Å and 4.308 Å with all <K-O-K angles as 90°. The observed value of K-K diagonal distance is 5.8471 Å. Four Na atoms form a parallelogram-shaped alkali cage that surrounds the dimers in orange (Dimer-II) and is schematically shown in Fig. 16(b). The Na-Na side distances in the cage are 4.308 Å and 5.8471 Å with <Na-Na-Na angles as 42.54° and 137.46°. The Na-Na diagonal distances are found as 3.9534 Å and 9.4797 Å. We also observed significant modifications in the <K-O-K (for Dimer-I) and <Na-O-Na (for Dimer-II) [see Table 1]. Therefore, compared to pure $NaO_2$, dramatic changes in the diagonal distances (enhancement) as well as in the <Na-O-Na angles (reduction or enhancement) are observed for $Na_{0.5}K_{0.5}O_2$. As a result, a pronounced structural deformation is observed in $Na_{0.5}K_{0.5}O_2$ that significantly reduces the crystal symmetry and hence presumably should lift the orbital degeneracy in the O-$\pi^*$ orbitals.

Below we describe the DOS of 50% doped one (the corresponding compound is $Na_{0.5}K_{0.5}O_2$) [see Fig. 17] and the other two cases are discussed in comparison. In comparison with LSDA DOS of monoclinic $KO_2$ [see Fig. 14], we find that $Na_{0.5}K_{0.5}O_2$ is half-metallic and has a similar pseudogap at $E_F$, unlike marcasite $NaO_2$ [see Fig. 7]. Comparing with marcasite $NaO_2$ [see Fig. 7], there are signatures of the lifting of orbital degeneracy among the $\pi^*$ orbitals around the $E_F$ upon doping. Since 4s electrons of K atom are more loosely bound to its ionic core than the corresponding 3s electrons of Na atoms, the electrostatic interaction between alkali ions and the $O_2^-$ dimers will be stronger in $KO_2$ than for $NaO_2$. Thus, the net electrostatic force is increased upon K doping. As a result, restriction in the movement of O atoms of $O_2^-$ dimers is stronger for those in the plane containing K ions compared to the plane containing Na ions. As consequence, the strength of the magnetogyration effect will increase with K doping at Na site in $NaO_2$. The lowering of symmetry in $NaO_2$ crystal with K doping concentration leads to the breaking of orbital degeneracy among the $\pi^*$ orbitals. This should lead to full OO and MIT in $Na_{0.5}K_{0.5}O_2$ at a lower value of U compared to pure $NaO_2$ in the same phase and possibly at a higher temperature than that in $KO_2$.

We found that $Na_{0.5}K_{0.5}O_2$ encounters an MIT at U=4 eV with a band gap of 0.3 eV. The TDOS and PDOS of $Na_{0.5}K_{0.5}O_2$ for U=0 and 4 eV are illustrated respectively in Fig. 17(a) and 17(b). However, we do not present here the electronic and magnetic properties for the other two levels (25% and 75%) of K doping for which $NaO_2$ also encounters MITs but comparatively at higher U values (5 eV for both $Na_{0.75}K_{0.25}O_2$ and $Na_{0.25}K_{0.75}O_2$). The associated magnetic properties of K-doped $NaO_2$ will be dealt with in future.

## 4. Conclusions

In summary, both marcasite $NaO_2$ and monoclinic $KO_2$ are found half-metallic in the LSDA calculations. The electrostatic interaction between the alkali ions and the $O_2^-$ dimers is qualitatively different between these two materials. This is so because the 4s electron of K atom is more delocalized than the 3s electrons of Na atom. The enhanced electrostatic interaction leads to larger distortion in the structure, namely the alkali cage is more deformed in $KO_2$ than in $NaO_2$. In the case of $KO_2$, we observed a pseudogap developing in the minority spin channel due to this orbital ordering, whereas there is no such effect apparently visible in more symmetric marcasite $NaO_2$. Furthermore, this is reflected in the lifting of degeneracy among the $\pi^*$ orbitals with smaller U in

$KO_2$, which is responsible for describing the electronic structure of these materials. We find in both cases correlation among the O-2p (π) electrons to be effective in completing OO and thereby driving the metal-insulator transition. The role of the alkali *s* electrons is found to be vital in raising the MIT temperature, as seen in the case of 25%, 50% and 75% K-doping levels in marcasite $NaO_2$. We found particularly in the case of $Na_{0.5}K_{0.5}O_2$ that a smaller correlation parameter (Hubbard U=4 eV) is successful in producing an MIT, highlighting the non-passive role of alkali *s* electrons in the electronic structure of superoxides. This is probably a way of obtaining high-temperature MIT.

**Acknowledgements**

The authors thank the West Bengal State University, Kolkata-700126, India and the Department of Higher Education, West Bengal, India for financial and infrastructure support to carry out this work.

# TABLE

**Table-1:** The Na-O bond lengths (Å) and <Na-O-Na bond angles (°) in the Na-octahedra around O-dimers of marcasite $NaO_2$, monoclinic $KO_2$ and $Na_{0.5}K_{0.5}O_2$. The O-O Dimer-I is surrounded by four K atoms and Dimer-II is surrounded by four Na atoms.

| Compounds | $NaO_2$ | $KO_2$ | $Na_{0.5}K_{0.5}O_2$ | |
|---|---|---|---|---|
| | | | Dimer-I | Dimer-II |
| **Bond lengths (Å)** | | | | |
| Na/K-O (apical) | 2.3814 (x2) | 2.8276 (x2) | 2.5539 (x2) | 2.0918 (x2) |
| Na/K-O (basal) | 2.4061 (x4) | 2.7829 (x2) | 2.2369 (x2) | 2.8971 (x2) |
| | | 2.8353 (x2) | 2.5539 (x2) | 2.0918 (x2) |
| O-O (dimer) | 1.31 | 1.3534 | 1.3285 | 1.3686 |
| **Bond angles** | | | | |
| <Na/K-O-Na/K (apical) | 108.88 | 108.77 | 107.921 | 118.479 |
| <Na/K-O-Na/K (basal) | 91.263 | 96.418 | 101.425 | 101.730 |
| | | 91.239 | | |

# FIGURE CAPTIONS

**Fig. 1:** Crystal structure of $NaO_2$ in the marcasite Pnnm phase (a). The formation of rhombus and rectangular Na cage in this phase is schematically illustrated in (b). The crystal structure of $NO_2$ in the pyrite Pa-3 phase (c) and schematic representation of the formation of square-shaped Na cage (d) in pyrite $NaO_2$.

**Fig. 2:** The total (a) and PBS of O-$p_{x/y/z}$ (b) orbitals of pyrite $NaO_2$ for the spin majority channel for U=0 eV. For the spin minority channel, the corresponding band structures are illustrated in (c) and (d) respectively. The zero value in the energy scale corresponds to $E_F$.

**Fig. 3:** The spin-polarized TDOS and PDOS of pyrite $NaO_2$ for U=0 eV. The zero value in the energy scale corresponds to $E_F$.

**Fig. 4:** The spin-polarized TDOS and PDOS of pyrite $NaO_2$ for U=6 eV (b). The zero value in the energy scale corresponds to $E_F$.

**Fig. 5:** The total (a) and partial band structure of O-$p_{x/y}$ (b), $p_z$ (c) orbitals of marcasite $NaO_2$ for the spin majority channel (1$^{st}$ panel) for U=0 eV. For the spin minority channel (2$^{nd}$ panel), the corresponding band structures are illustrated in (d), (e), (f). The zero value in the energy scale corresponds to $E_F$.

**Fig. 6:** Schematic depiction of possible splitting of O-p states in marcasite phase of $NaO_2$ for U=0 eV. Each of the σ, π$_+$ and π$_-$ states are occupied by two electrons and the π$_+^*$ and π$_-^*$ states are occupied by one electron each. The remaining electron donated by Na atom has an equal probability to occupy the π$_+^*$, π$_-^*$ and σ$^*$ states. The O-2p electrons are represented by red and Na-3s electrons by the blue arrow.

**Fig. 7:** The spin-polarized TDOS (black) and PDOS of Na-s (blue), Na-p (green) and O-p (red) orbitals of marcasite $NaO_2$ for U=0 eV. The zero value in the energy scale corresponds to $E_F$.

**Fig. 8:** The spin-polarized TDOS of marcasite $NaO_2$ for different U values starting from 3 eV to 6 eV. The system remains half-metallic up to U=5 eV and encounters MIT at U = 6 eV with a band gap of 0.3 eV. The zero value in the energy scale corresponds to $E_F$.

**Fig. 9:** The total (a) and partial band structure of O-p$_{x/y}$ (b), p$_z$ (c) orbitals of marcasite NaO$_2$ for the spin majority channel (1$^{st}$ panel) at U=6 eV. In the spin minority channel (2$^{nd}$ panel), the corresponding band structures are illustrated in (d), (e), (f). The zero value in the energy scale corresponds to E$_F$.

**Fig. 10:** Schematic depiction of possible splitting of O-p states in marcasite phase of marcasite NaO$_2$ for U=6 eV. All of the σ, π$_+$ and π. states are occupied by two electrons each. The π$_+^*$ and π.$^*$ states are occupied respectively by two and one electrons. The extra unpaired electron donated by the Na atom is stabilized now, leading to full range orbital ordering in the O-2p states. The O-2p electrons are represented by red and Na-3s electrons by the blue arrow.

**Fig. 11:** The spin-polarized TDOS (black) and PDOS of Na-s (blue), Na-p (green) and O-p (red) orbitals of marcasite NaO$_2$ for U=6 eV. The zero value in the energy scale corresponds to E$_F$.

**Fig. 12:** Crystal structure of KO$_2$ in the monoclinic C2/c phase (a). The formation of parallelogram-shaped K cage in this phase is schematically illustrated in (b). The crystal structure of KO$_2$ in the tetragonal I4/mmm phase (c), schematic representation of the formation of square-shaped K cage (d) tetragonal KO$_2$ and the crystal structure of Na$_{0.5}$K$_{0.5}$O$_2$ for the monoclinic phase (e).

**Fig. 13:** The total (a) and partial band structure of O-p$_x$ (b), p$_y$ (c), p$_z$ (d) orbitals of monoclinic KO$_2$ for the spin majority channel (1$^{st}$ panel) are illustrated for U=0 eV. In the spin minority channel (2$^{nd}$ panel), the corresponding band structures are illustrated in (e), (f), (g), (h). The zero value in the energy scale corresponds to E$_F$.

**Fig. 14:** The spin-polarized TDOS (black) and PDOS of K-s (blue), K-p (green) and O-p (red) orbitals of monoclinic KO$_2$ for U=0 eV. The zero value in the energy scale corresponds to E$_F$.

**Fig. 15:** The spin-polarized TDOS (black) and PDOS of K-s (blue), K-p (green) and O-p (red) orbitals of monoclinic KO$_2$ for U=4 eV. A band gap of 0.3 eV is opened near the E$_F$. The zero value in the energy scale corresponds to E$_F$.

**Fig. 16:** Crystal structure of $Na_{0.5}K_{0.5}O_2$ (a). The formation of two kinds of alkali cages are schematically shown in (b). The alkali cages surrounding the red-coloured dimers (Dimer-I) are rectangular-shaped and the alkali cages surrounding the orange-coloured dimers (Dimer-II) are parallelogram-shaped (b).

**Fig. 17:** The spin-polarized TDOS (black) and PDOS of Na-p (blue), Na-s (green), K-p (orange), K-s (violet) and O-2p (red) orbitals of $Na_{0.5}K_{0.5}O_2$ for U=0 (a) and U=4 eV (b). The system encounters MIT with a band gap of $E_g$ = 0.3 eV. The zero value in the energy scale corresponds to $E_F$.

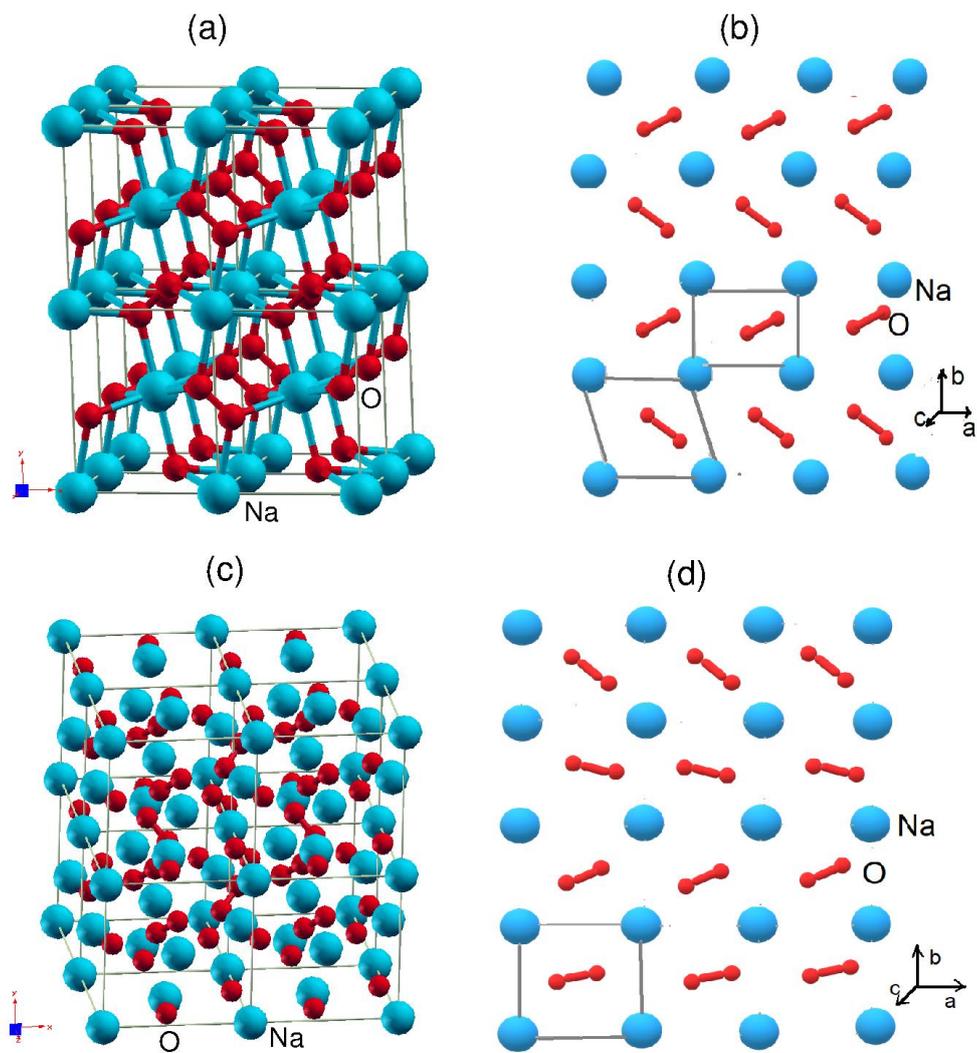

**Fig. 1**

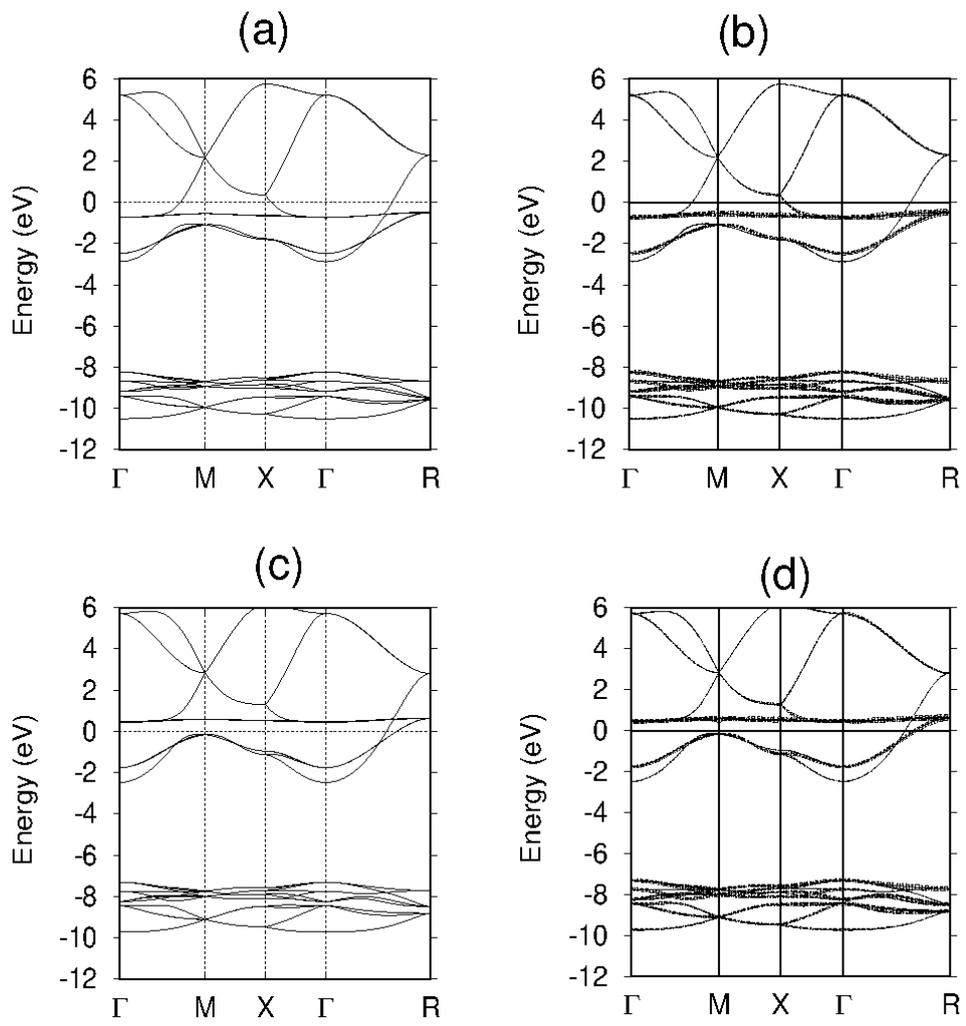

**Fig. 2**

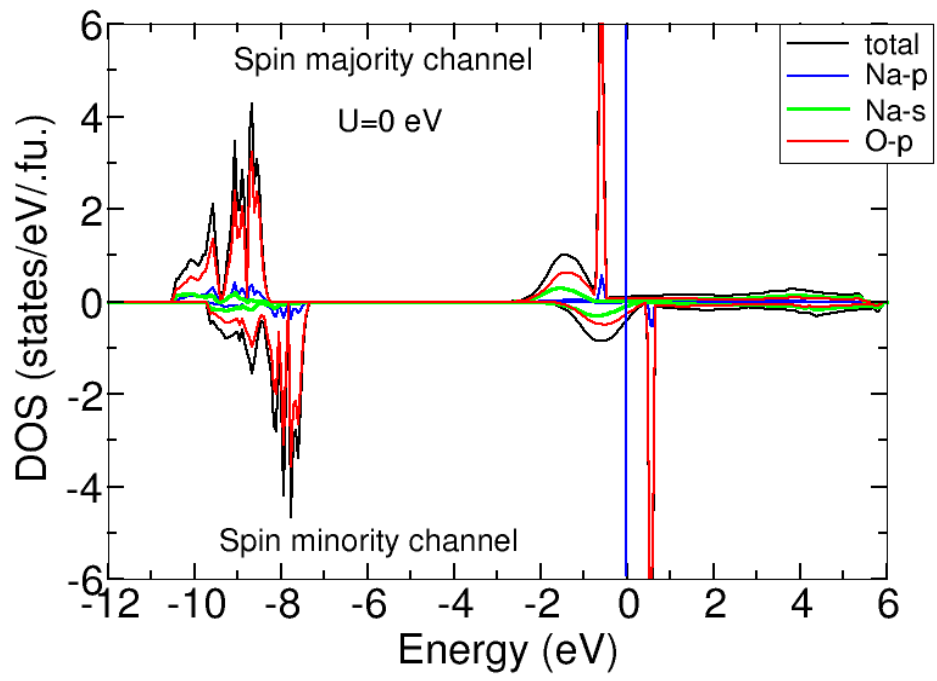

**Fig. 3**

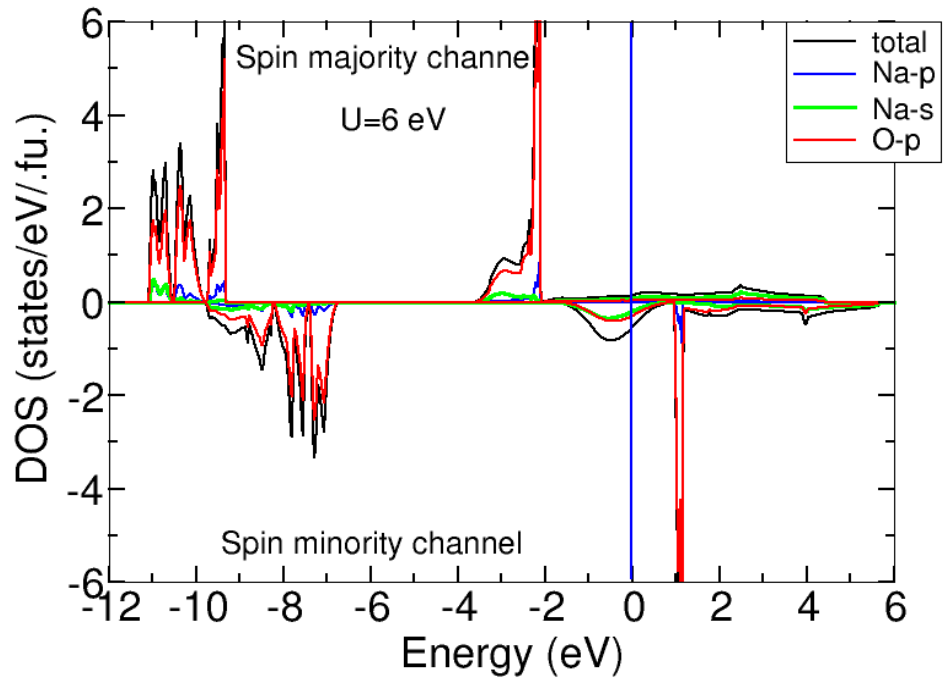

**Fig. 4**

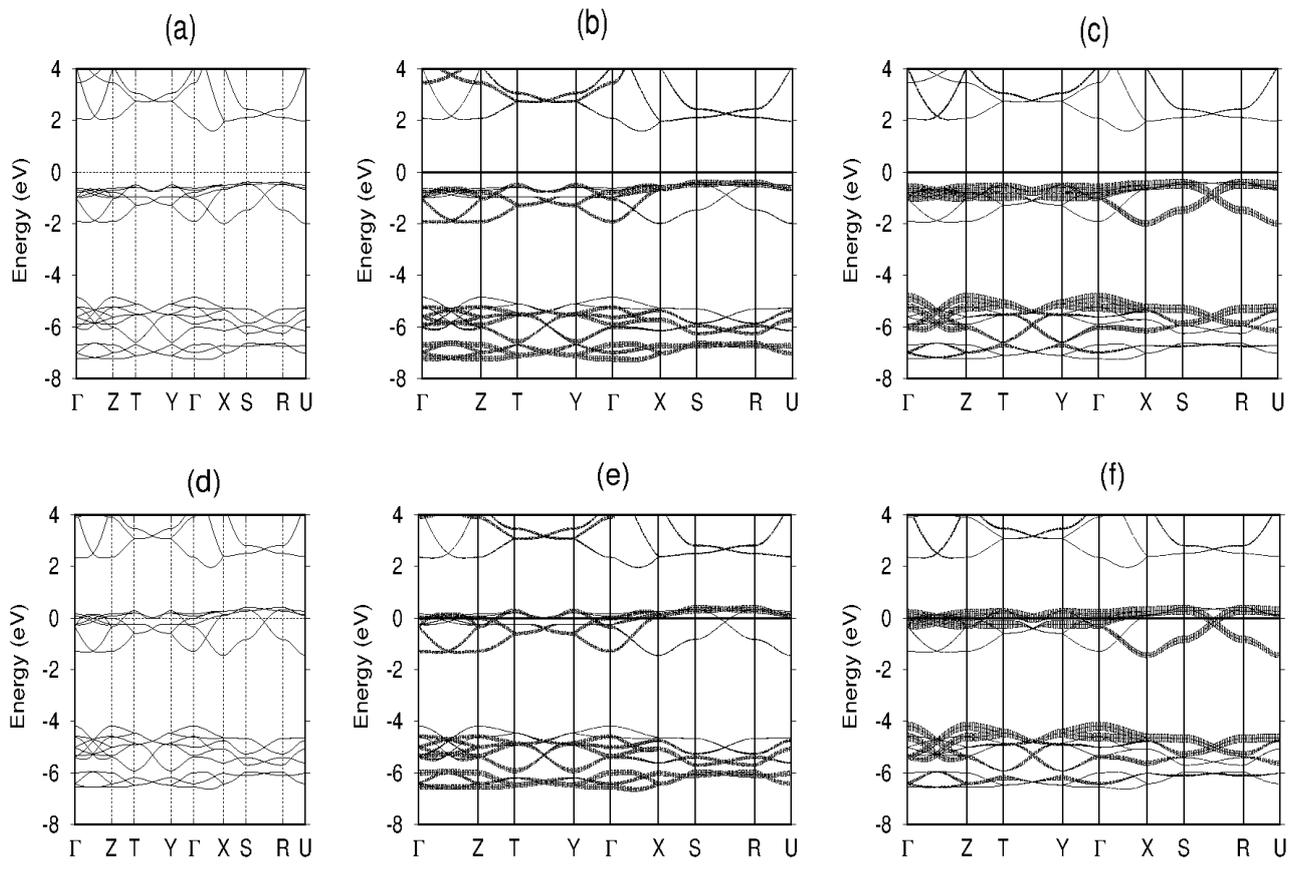

**Fig. 5**

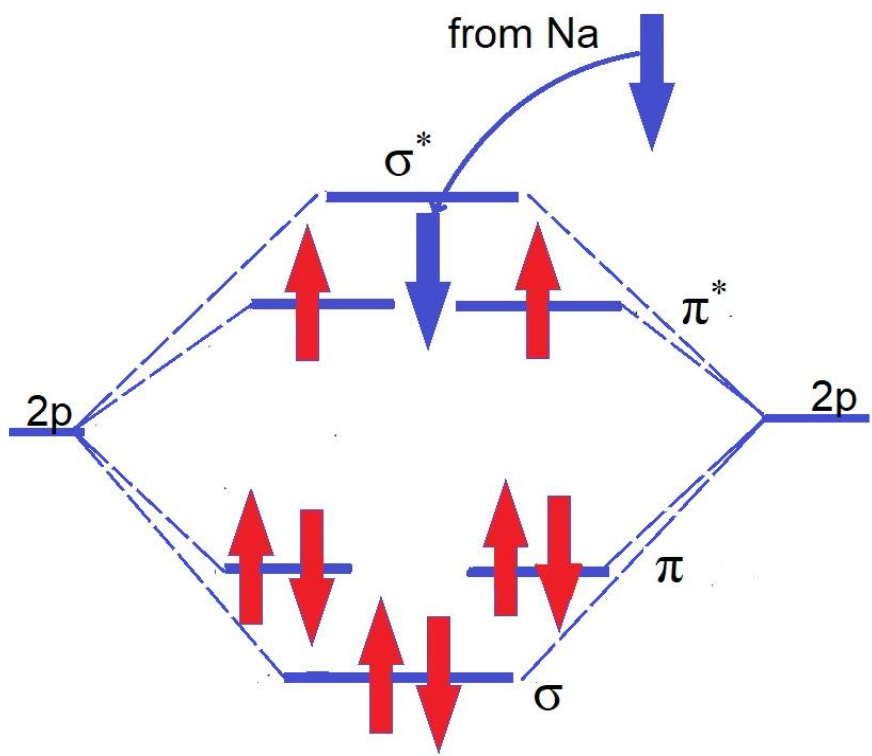

**Fig. 6**

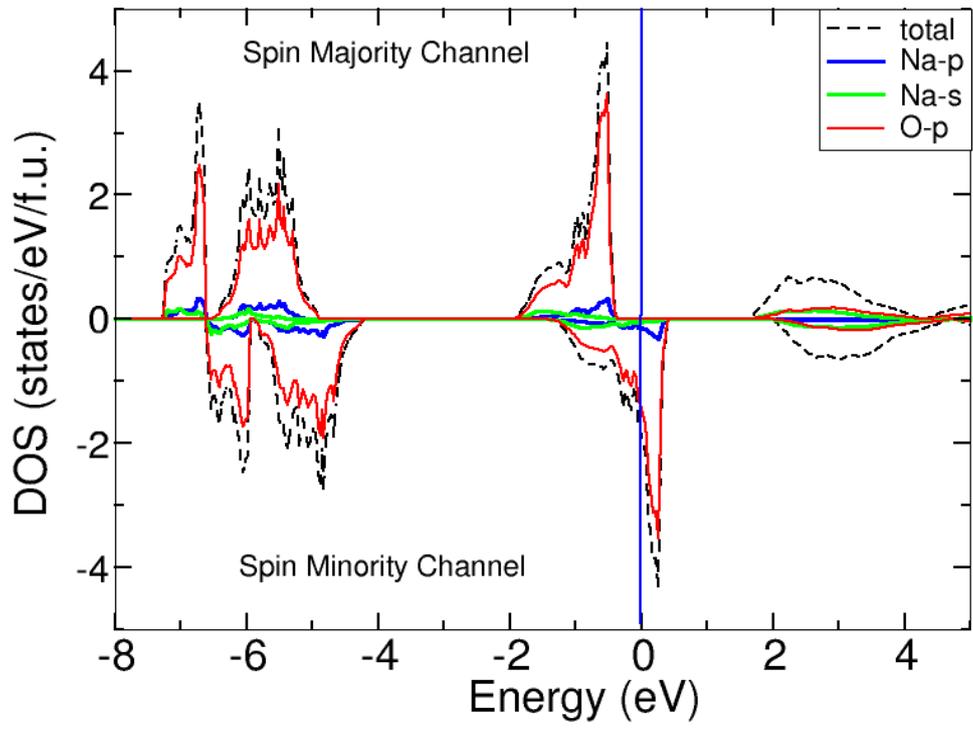

**Fig. 7**

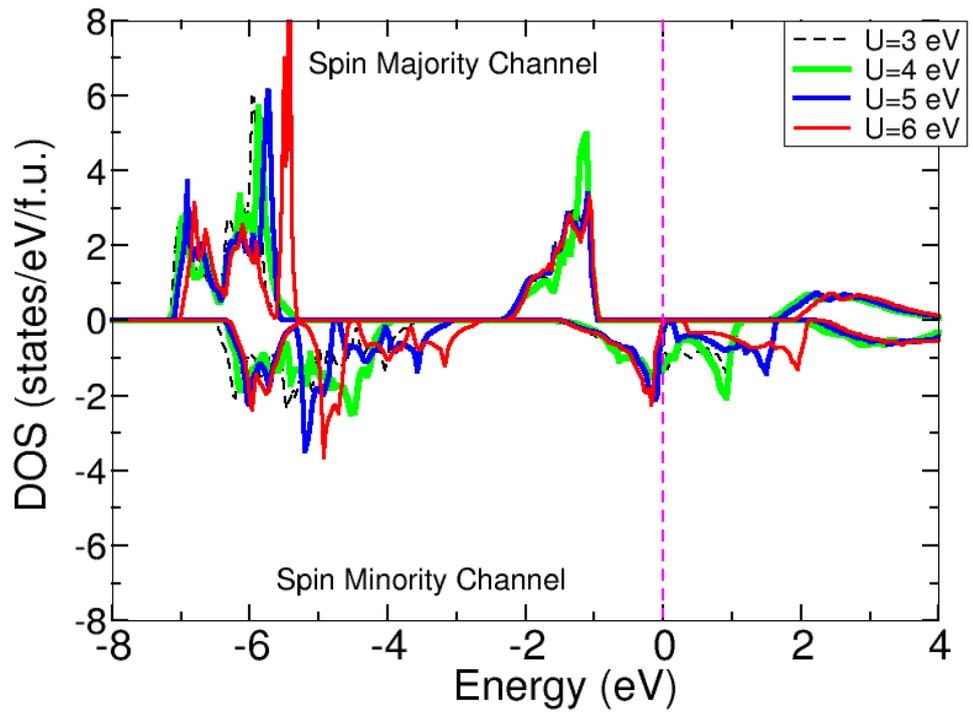

**Fig. 8**

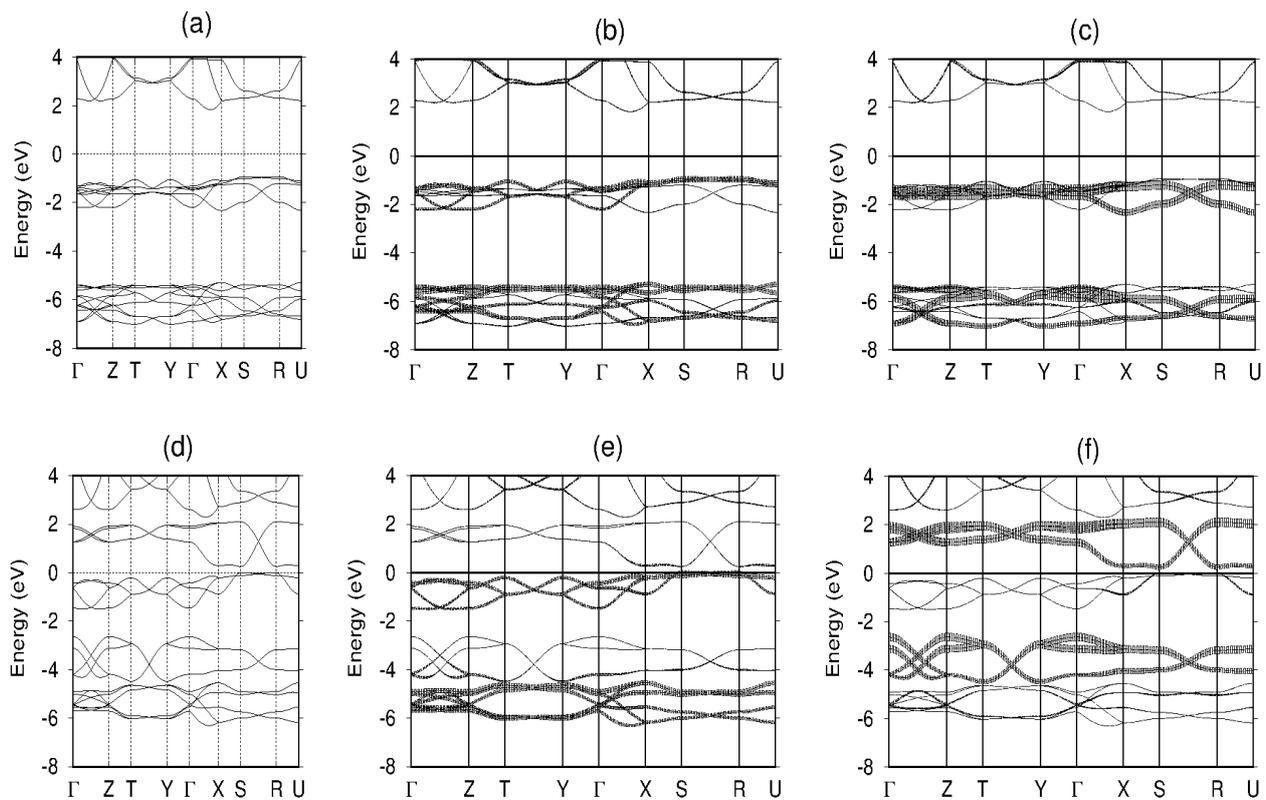

**Fig. 9**

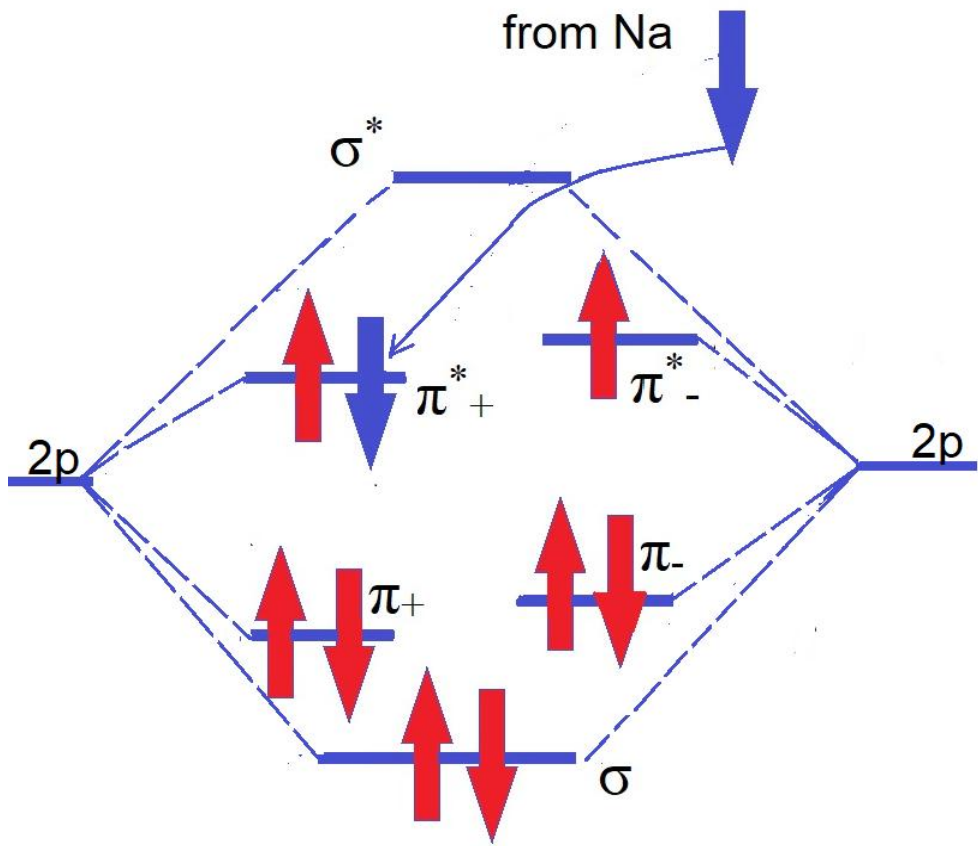

**Fig. 10**

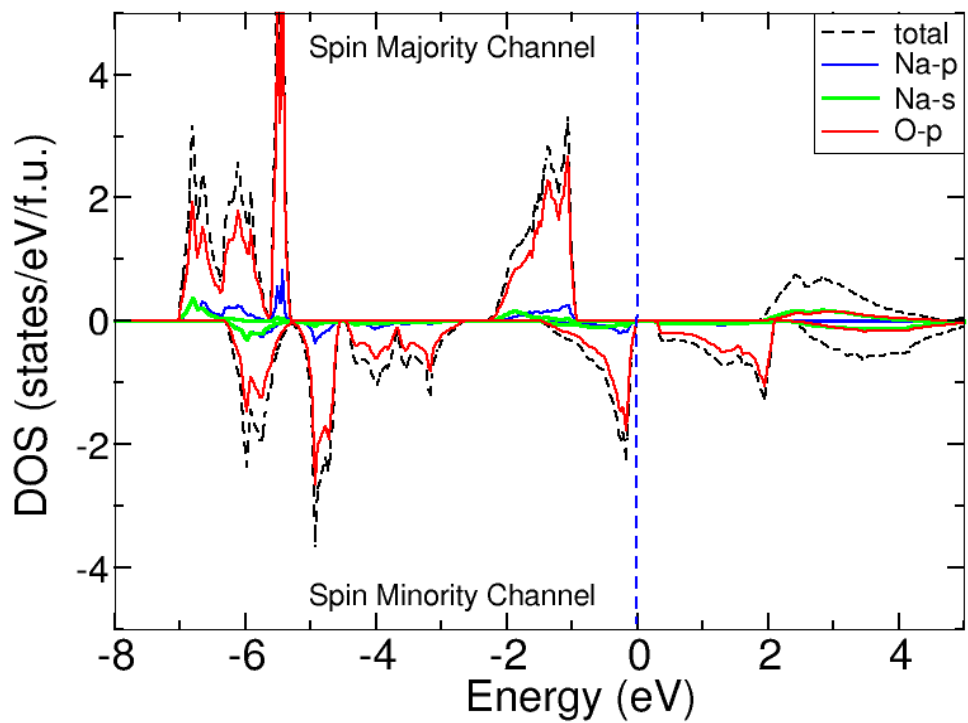

**Fig. 11**

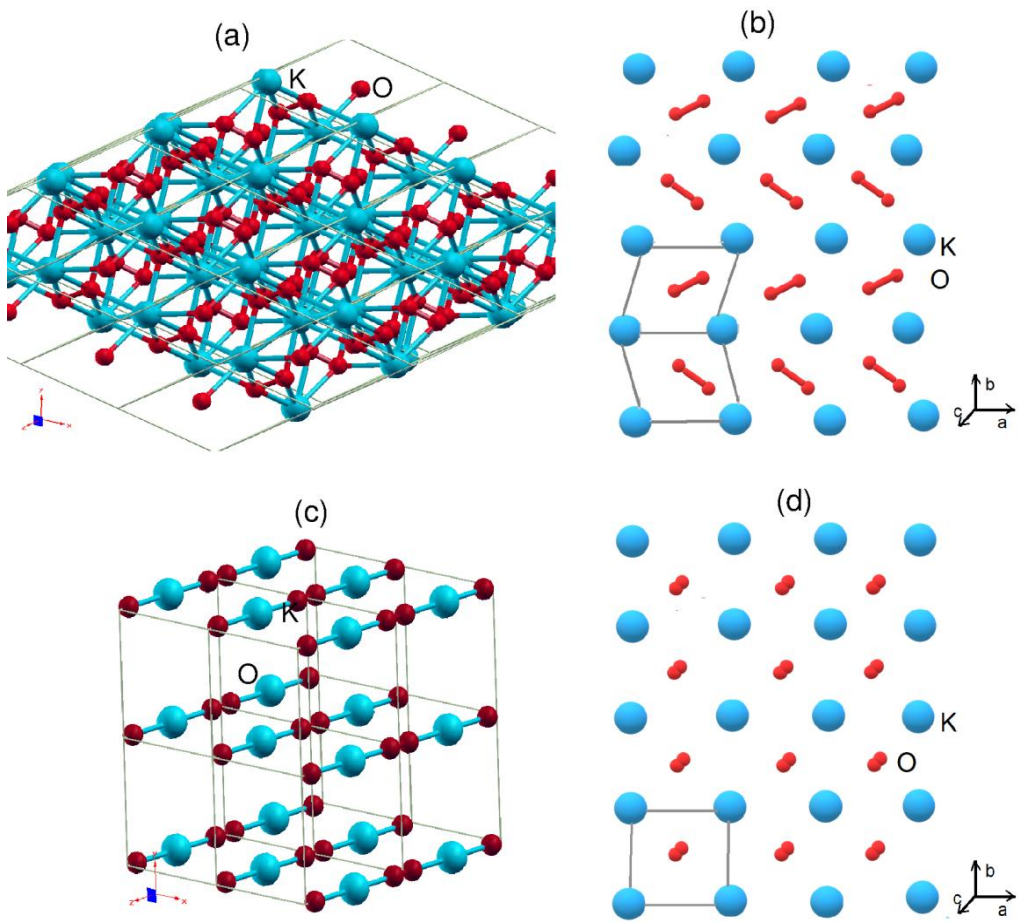

**Fig. 12**

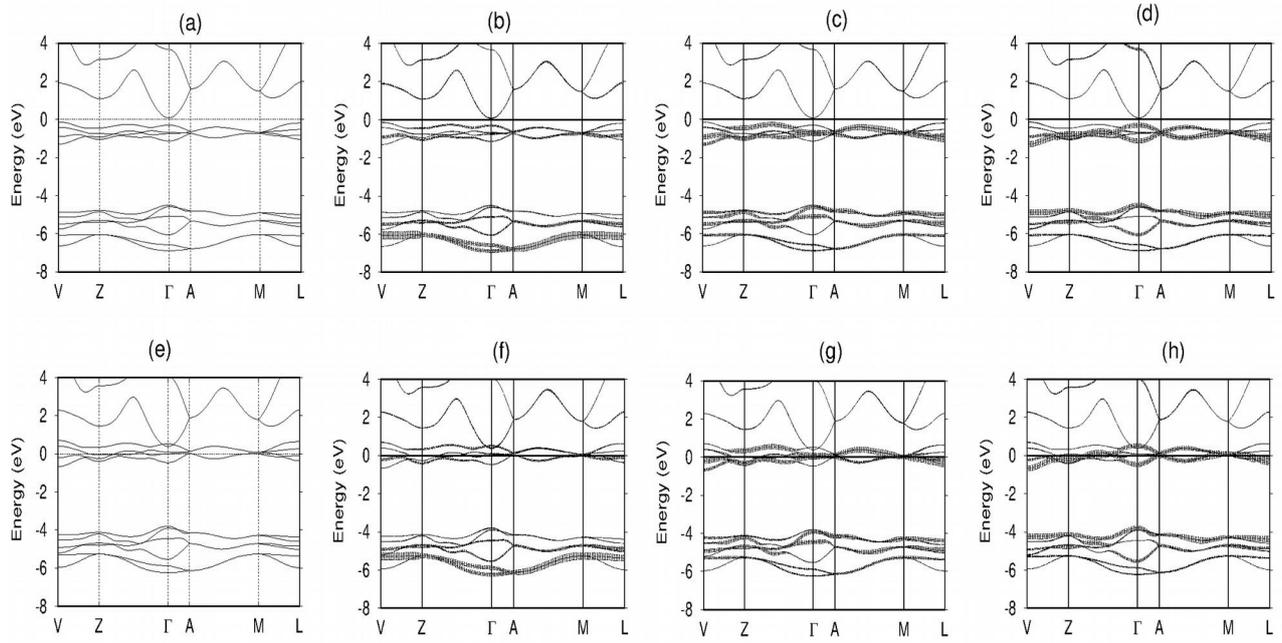

**Fig. 13**

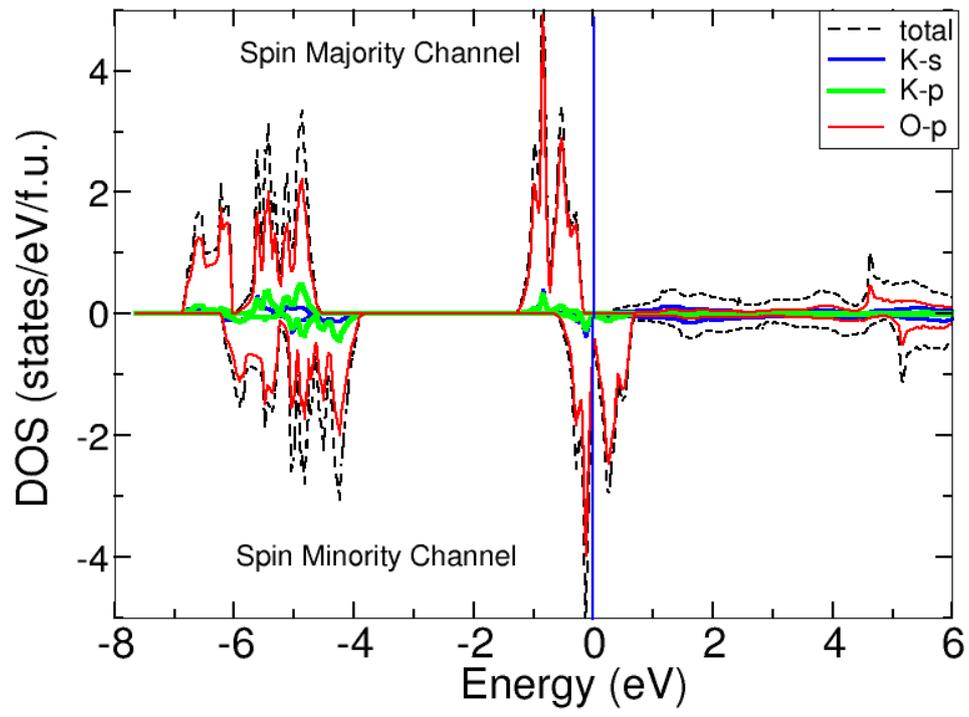

**Fig. 14**

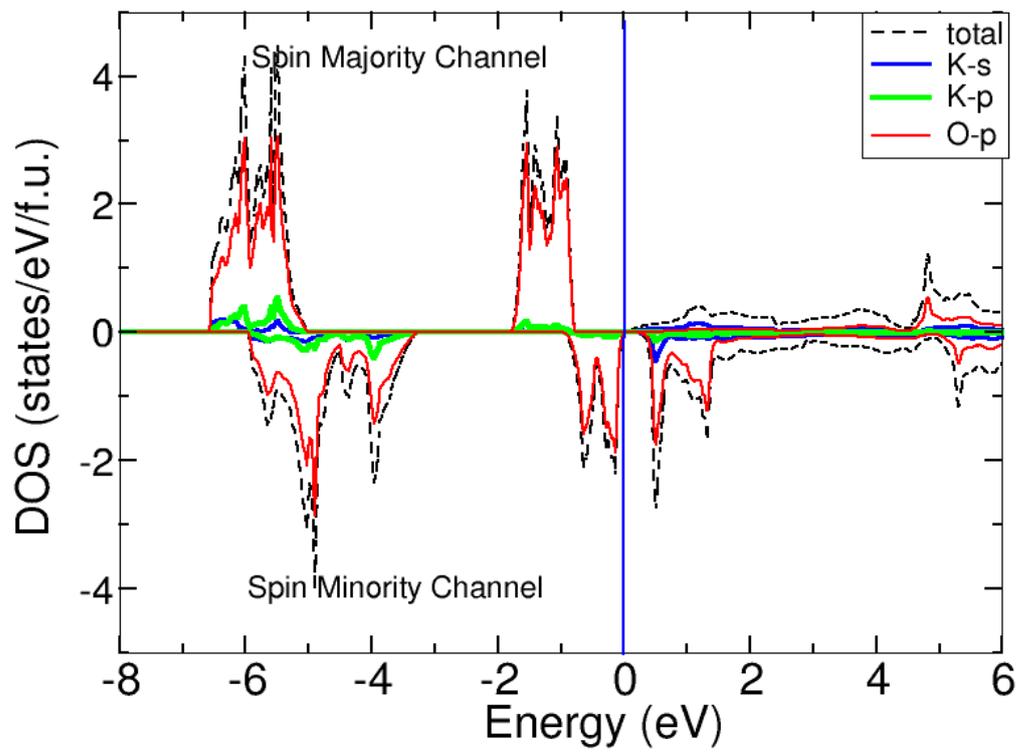

**Fig. 15**

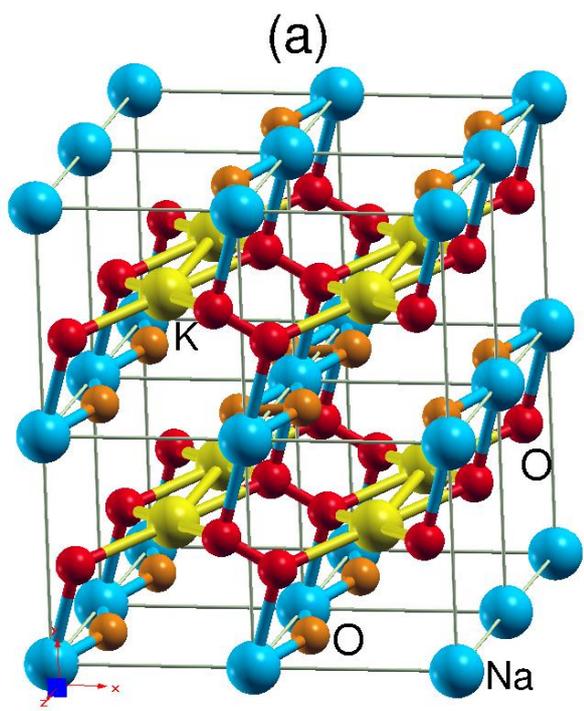 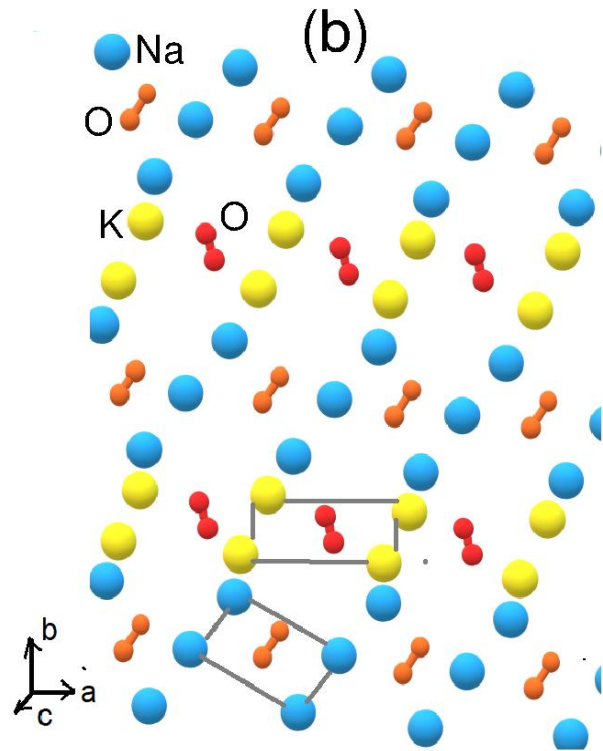

**Fig. 16**

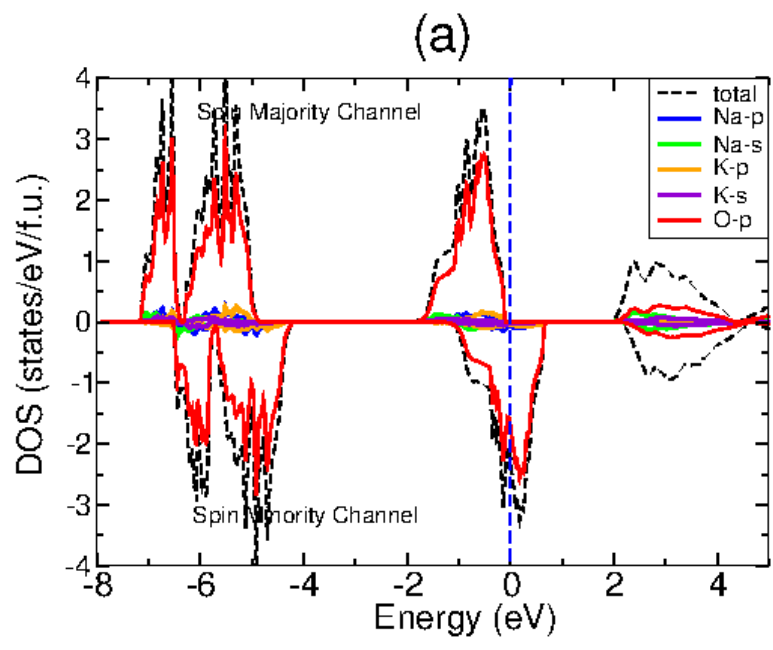

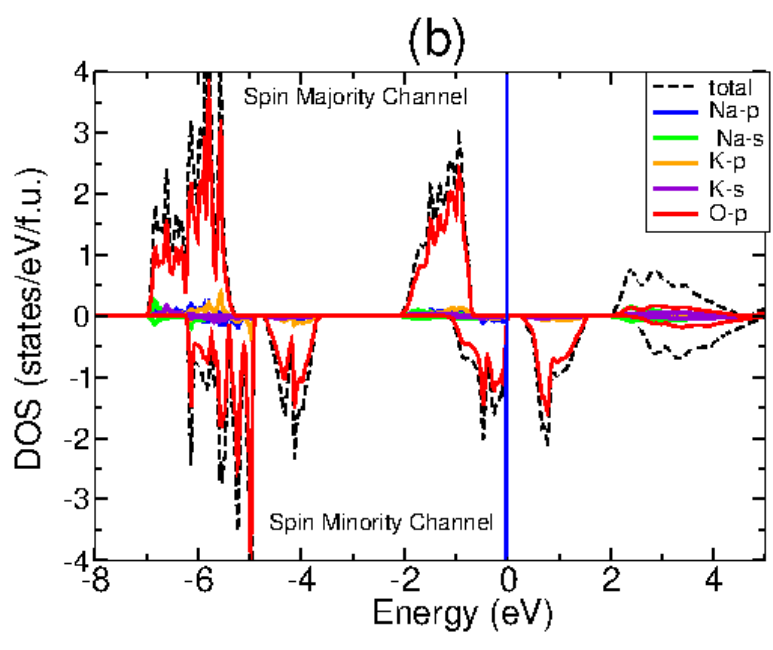

**Fig. 17**